\documentstyle[12pt]{article}
\newcommand{\Cslash}{\not \!\! C}
\newcommand{\Dslash}{\not \!\! D}
\newcommand{\Aslash}{\not \!\! A}
\newcommand{\kslash}{\not \!\! k}
\newcommand{\sslash}{\not \!\! s}
\newcommand{\pslash}{\not \!\! p}
\newcommand{\delslash}{\not \! \partial}
\begin{document}

\begin{flushright}{UT-02-32}
\end{flushright}
\vskip 0.5 truecm

\begin{center}
{\Large{\bf Generalized Ginsparg-Wilson algebra and index 
theorem on the lattice }}
\end{center}
\vskip .5 truecm
\centerline{\bf Kazuo Fujikawa}
\vskip .4 truecm
\centerline {\it Department of Physics,University of Tokyo}
\centerline {\it Bunkyo-ku,Tokyo 113,Japan}
\vskip 0.5 truecm

\makeatletter
\@addtoreset{equation}{section}
\def\theequation{\thesection.\arabic{equation}}
\makeatother

\begin{abstract}
Recent studies of  the topological properties of 
a general class of lattice Dirac operators are reported.
This is based on a  specific algebraic realization of the 
Ginsparg-Wilson relation in the form
 $\gamma_{5}(\gamma_{5}D)+(\gamma_{5}D)\gamma_{5} =
 2a^{2k+1}(\gamma_{5}D)^{2k+2}$ 
where $k$ stands for a non-negative integer. The choice  
$k=0$ corresponds to the commonly discussed Ginsparg-Wilson 
relation and thus to the overlap operator. 
It is shown that local chiral anomaly and the instanton-related 
index of all these operators are identical. The locality of all 
these Dirac operators for vanishing gauge fields is proved on 
the basis of explicit construction, but the locality with 
dynamical gauge fields has not been established yet. We suggest 
that the Wilsonian effective action is essential to avoid 
infrared singularities encountered in general perturbative 
analyses.
\end{abstract}

\section{Introduction}

Recent developments in the treatment of  fermions in lattice 
gauge theory are based on a  hermitian lattice Dirac  operator 
$\gamma_{5}D$ which satisfies the Ginsparg-Wilson 
relation\cite{ginsparg}
\begin{equation}
\gamma_{5}D + D\gamma_{5} = 2aD\gamma_{5}D
\end{equation}
where the lattice spacing $a$ is utilized to make a dimensional
consideration transparent, and 
$\gamma_{5}$ is a hermitian chiral Dirac matrix. 
An explicit example of the operator satisfying (1.1) and free of 
species doubling has been given by Neuberger.\cite{neuberger}  
The relation 
(1.1) led to an interesting analysis of the notion of index in 
lattice gauge theory.\cite{hasenfratz} This index theorem in 
turn led to a 
new form of chiral symmetry, and the chiral anomaly is obtained 
as a non-trivial Jacobian factor under this modified chiral
 transformation.\cite{luscher} This chiral Jacobian is regarded 
as a lattice 
generalization of the continuum path integral.\cite{fujikawa} 
The very 
detailed analyses of the lattice chiral Jacobian have been 
performed.\cite{kikukawa} It is also possible to formulate the 
lattice 
index theorem in a manner analogous to the continuum index 
theorem.\cite{jackiw}${}^{,}$\cite{atiyah}${}^{,}$\cite{adler} 
An interesting 
chirality sum rule, which 
relates the number of zero modes to that of the heaviest states, 
has also been noticed.\cite{chiu} See Refs.\cite{niedermayer} 
for reviews of these
developments. 

We have recently discussed the possible generalization of (1.1)
and its implications.\cite{fujikawa2} To be specific, we have
discussed a generalization of the algebra (1.1) in the form 
\begin{equation}
\gamma_{5}(\gamma_{5}D)+(\gamma_{5}D)\gamma_{5}=
2a^{2k+1}(\gamma_{5}D)^{2k+2}
\end{equation}
where $k$ stands for a non-negative integer and $k=0$ corresponds
to the ordinary Ginsparg-Wilson relation. 
 When one defines 
\begin{equation}
H\equiv \gamma_{5}aD
\end{equation}
(1.2) is rewritten as 
\begin{equation}
\gamma_{5}H+H\gamma_{5}=2H^{2k+2}
\end{equation}
or equivalently
\begin{equation}
\Gamma_{5}H+H\Gamma_{5}=0
\end{equation}
where we defined 
\begin{equation}
\Gamma_{5}\equiv \gamma_{5}-H^{2k+1}.
\end{equation}
Note that both of $H$ and $\Gamma_{5}$ are hermitian operators.

It has been shown that all the good topological properties of 
the overlap
operator\cite{neuberger} is retained in this 
generalization.\cite{fujikawa2,chiu2} The 
practical 
applications of this generalization are not known at this moment.
We however mention the charactristic properties of this 
generalization: The spectrum near the continuum configuration 
is closer to that of continuum theory and the chiral symmetry 
breaking terms become more irrelevent in the continuum limit
for $k\geq 1$. The operator however spreads over more lattice
points for large $k$.

\section{Representation of the general algebra}

We first discuss a general representation of the algebraic 
relation (1.5). 
The relation (1.5) suggests that if
\begin{equation}
H\phi_{n} = a\lambda_{n}\phi_{n}, \ \ \ (\phi_{n},\phi_{n})=1 
\end{equation}
with a real eigenvalue $a\lambda_{n}$ for the hermitian 
operator $H$, then
\begin{equation}
H(\Gamma_{5}\phi_{n}) = -a\lambda_{n}(\Gamma_{5}\phi_{n}).
\end{equation}
Namely, the eigenvalues $\lambda_{n}$ and $-\lambda_{n}$ are 
always paired if $\lambda_{n}\neq 0$ and 
$(\Gamma_{5}\phi_{n},\Gamma_{5}\phi_{n})\neq 0$.
We also note the relation, which is derived by sandwiching the
relation (1.4) by $\phi_{n}$,
\begin{equation}
(\phi_{n},\gamma_{5}\phi_{n})=(a\lambda_{n})^{2k+1}\ \ \ \ for\ \
 \ \lambda_{n}\neq 0.
\end{equation}
Consequently
\begin{equation}
|(a\lambda_{n})^{2k+1}|= |(\phi_{n},\gamma_{5}\phi_{n})|\leq
||\phi_{n}||||\gamma_{5}\phi_{n}||=1.
\end{equation}
Namely, all the possible eigenvalues are bounded by
\begin{equation}
|\lambda_{n}|\leq \frac{1}{a}.
\end{equation}

We thus  evaluate the norm of $\Gamma_{5}\phi_{n}$
\begin{eqnarray}
(\Gamma_{5}\phi_{n},\Gamma_{5}\phi_{n})
&=&(\phi_{n},(\gamma_{5}-H^{2k+1})(\gamma_{5}-H^{2k+1})\phi_{n})
\nonumber\\ 
&=&
(\phi_{n},(1-H^{2k+1}\gamma_{5}-\gamma_{5}H^{2k+1}+H^{2(2k+1)})
\phi_{n})\nonumber\\
\end{eqnarray}
where we used (2.3).
By remembering that all the eigenvalues are real, we find that 
$\phi_{n}$ is a ``highest'' state 
\begin{equation}
\Gamma_{5}\phi_{n}=0
\end{equation}
only if 
\begin{equation}
[1-(a\lambda_{n})^{2}]=(1-a\lambda_{n})(1+a\lambda_{n})=0 
\end{equation}
for the Euclidean positive definite inner 
product $(\phi_{n}, \phi_{n})\equiv\sum_{x}\phi_{n}^{\dagger}(x)
\phi_{n}(x)$.\\
We thus conclude that the states $\phi_{n}$ with 
$\lambda_{n}= \pm \frac{1}{a}$
 are {\em not} paired by the operation $\Gamma_{5}\phi_{n}$ and 
\begin{equation}
\gamma_{5}D\phi_{n}= \pm \frac{1}{a}\phi_{n}, \ \ \ \gamma_{5}
\phi_{n}= \pm \phi_{n}
\end{equation}
respectively. These eigenvalues are in fact the maximum or 
minimum of the possible eigenvalues of $H/a$ due to (2.5).

As for the vanishing eigenvalues $H\phi_{n}=0$, we find from
(1.4) that $H\gamma_{5}\phi_{n}=0$, namely, 
$H[(1\pm\gamma_{5})/2]\phi_{n}=0$. We can thus choose
\begin{equation}
\gamma_{5}D\phi_{n}=0,\ \ \ \gamma_{5}\phi_{n}=\phi_{n} \ \ \ or \ \ \ 
\gamma_{5}\phi_{n}=-\phi_{n}.
\end{equation}
\\
To summarize the analyses so far, all the normalizable 
eigenstates $\phi_{n}$ of $\gamma_{5}D=H/a$ are categorized into 
the following 3 classes:\\
(i)\ $n_{\pm}$ (``zero modes''),\\
\begin{equation}
\gamma_{5}D\phi_{n}=0, \ \ \gamma_{5}\phi_{n} = \pm \phi_{n},
\end{equation}
(ii)\ $N_{\pm}$ (``highest states''), \\
\begin{equation}
\gamma_{5}D\phi_{n}= \pm \frac{1}{a}\phi_{n}, \ \ \
\gamma_{5}\phi_{n} = \pm \phi_{n},\ \ \ respectively,
\end{equation}
(iii)``paired states'' with $0 < |\lambda_{n}| < 1/a$,
\begin{equation}
\gamma_{5}D\phi_{n}= \lambda_{n}\phi_{n}, \ \ \ 
\gamma_{5}D(\Gamma_{5}\phi_{n})
= - \lambda_{n}(\Gamma_{5}\phi_{n}).
\end{equation}
Note that $\Gamma_{5}(\Gamma_{5}\phi_{n})\propto \phi_{n}$ for 
$0<|\lambda_{n}|<1/a$.\\

We thus obtain the index relation\cite{hasenfratz,luscher}
\begin{eqnarray}
Tr\Gamma_{5}&\equiv& \sum_{n}(\phi_{n},\Gamma_{5}\phi_{n})
\nonumber\\
&=&\sum_{ \lambda_{n}=0}(\phi_{n},\Gamma_{5}\phi_{n})
+\sum_{0<|\lambda_{n}|<1/a}(\phi_{n},\Gamma_{5}\phi_{n})
+\sum_{|\lambda_{n}|=1/a}(\phi_{n},\Gamma_{5}\phi_{n})
\nonumber\\
&=&\sum_{\lambda_{n}=0}(\phi_{n},\Gamma_{5}\phi_{n})\nonumber\\
&=&\sum_{\lambda_{n}=0}(\phi_{n},(\gamma_{5}-H^{2k+1})\phi_{n})
\nonumber\\
&=&\sum_{\lambda_{n}=0}(\phi_{n},\gamma_{5}\phi_{n})
\nonumber\\
&=& n_{+} - n_{-} =  index
\end{eqnarray}
where $n_{\pm}$ stand for the number of  normalizable zero modes
with $\gamma_{5}\phi_{n}=\pm\phi_{n}$ in the classification (i) 
above. We here used the fact that 
$\Gamma_{5}\phi_{n}=0$ for the ``highest states'' and that 
$\phi_{n}$ and $\Gamma_{5}\phi_{n}$ are orthogonal to each other
for $0<|\lambda_{n}|<1/a$ since they have eigenvalues 
with opposite signatures.

On the other hand, the relation $Tr \gamma_{5}=0$, which is
expected to be valid in (finite) lattice theory,  leads to ( by 
using (2.3))
\begin{eqnarray}
Tr\gamma_{5} &=& \sum_{n}(\phi_{n},\gamma_{5}\phi_{n})\nonumber\\
&=& \sum_{ \lambda_{n}=0}(\phi_{n},\gamma_{5}\phi_{n}) +
\sum_{\lambda_{n}\neq 0}(\phi_{n},\gamma_{5}\phi_{n})\nonumber\\
&=& n_{+} - n_{-}+\sum_{ \lambda_{n}\neq 0}(a\lambda_{n})^{2k+1}
=0.
\end{eqnarray}
In the last line  of this relation, all the states except for 
the 
``highest states'' with $\lambda_{n}= \pm 1/a$  cancel pairwise 
for $\lambda_{n}\neq0$. We thus obtain a chirality  sum 
rule[10]  
\begin{equation}
n_{+}+N_{+}=n_{-}+N_{-}  
\end{equation} 
where $N_{\pm}$ stand for the number of ``highest states'' with
$\gamma_{5}\phi_{n}=\pm\phi_{n}$ in the classification (ii) 
above. These 
relations show that the chirality asymmetry at vanishing 
eigenvalues is balanced by the chirality asymmetry at the 
largest eigenvalues with $|\lambda_{n}|=1/a$. It was argued in 
Ref.[14] that $N_{\pm}$ states are the topological 
(instanton-related) excitations of the would-be species doublers.

We have thus established that the representation of all the 
algebraic relations (1.2) has a similar structure. In the next
section, we show that the index $n_{+}-n_{-}$ is identical
to all these algebraic relations if the operator $\gamma_{5}D$
satisfies suitable conditions.
 
\section{Chiral Jacobian and the index relation}

The Euclidean path integral for a fermion is defined by
\begin{equation}
\int{\cal D}\bar{\psi}{\cal D}\psi\exp[\int\bar{\psi}D\psi]
\end{equation}
where
\begin{equation}
\int\bar{\psi}D\psi\equiv \sum_{x,y}\bar{\psi}(x)D(x,y)\psi(y)
\end{equation}
and the summation runs over all the points on the lattice.
The relation (1.5) is re-written as 
\begin{equation}
\gamma_{5}\Gamma_{5}\gamma_{5}D+D\Gamma_{5}=0
\end{equation}
and thus the Euclidean action is invariant under the global
 ``chiral'' transformation\cite{luscher}
\begin{eqnarray}
&&\bar{\psi}(x)\rightarrow\bar{\psi}^{\prime}(x)=
\bar{\psi}(x)+i\sum_{z}\bar{\psi}(z)\epsilon\gamma_{5}
\Gamma_{5}(z,x)\gamma_{5}
\nonumber\\
&&\psi(y)\rightarrow\psi^{\prime}(y)=
\psi(y)+i\sum_{w}\epsilon\Gamma_{5}(y,w)\psi(w)
\end{eqnarray}
with an infinitesimal constant parameter $\epsilon$.
Under this transformation, one obtains a Jacobian factor
\begin{equation}
{\cal D}\bar{\psi}^{\prime}{\cal D}\psi^{\prime}=
J{\cal D}\bar{\psi}{\cal D}\psi
\end{equation}
with
\begin{equation}
J=\exp[-2iTr\epsilon\Gamma_{5}]=\exp[-2i\epsilon(n_{+}-n_{-})]
\end{equation}
where we used the index relation (2.14).

We now relate this index appearing in the Jacobian to the 
Pontryagin index of the gauge field in a smooth continuum limit 
by following the procedure in Ref.\cite{fujikawa2}.
We  start with
\begin{equation}
Tr\{\Gamma_{5}f(\frac{(\gamma_{5}D)^{2}}{M^{2}})\}
=Tr\{\Gamma_{5}f(\frac{(H/a)^{2}}{M^{2}})\}
=n_{+} - n_{-}
\end{equation}
Namely, the index is not modified by any  regulator $f(x)$ with 
$f(0)=1$ and $f(x)$ rapidly going to zero for 
$x\rightarrow\infty$, as can be confirmed by using (2.14). This
means that you can use {\em any} suitable $f(x)$ in the 
evaluation of the index by taking advantage of this property.

We then consider a local version of the index
\begin{equation}
tr\{\Gamma_{5}f(\frac{(\gamma_{5}D)^{2}}{M^{2}})\}(x,x)
=tr\{(\gamma_{5}-H^{2k+1})f(\frac{(\gamma_{5}D)^{2}}{M^{2}})\}
(x,x)
\end{equation}
where trace stands for Dirac and Yang-Mills indices; Tr in (3.7) 
includes a sum over the lattice points $x$.  
A local version of the index is not sensitive to the precise 
boundary condition , and  one may take an infinite volume 
limit of the lattice in the above expression. 

We now examine the continuum limit $a\rightarrow 0$ of the above 
local expression (3.8)\footnote{This continuum limit corresponds to the 
so-called ``naive'' continuum limit in the context of lattice gauge 
theory.}. We first observe that the term
\begin{equation}
tr\{H^{2k+1}f(\frac{(\gamma_{5}D)^{2}}{M^{2}})\}
\end{equation}
goes to zero in this limit. The large eigenvalues of 
$H=a\gamma_{5}D$ are truncated at the value $\sim aM$ by the 
regulator $f(x)$ which rapidly goes to zero for large $x$. In 
other words, the global index of the operator 
$TrH^{2k+1}f(\frac{(\gamma_{5}D)^{2}}{M^{2}})\sim O(aM)^{2k+1}
\rightarrow 0$ for $a\rightarrow 0$ with fixed $M$.

We thus examine the small $a$ limit of 
\begin{equation}
tr\{\gamma_{5}f(\frac{(\gamma_{5}D)^{2}}{M^{2}})\}.
\end{equation}
The operator appearing in this expression is well regularized by 
the function $f(x)$, and we evaluate the above trace by using 
the plane wave basis to extract an explicit gauge field 
dependence.
We consider a square lattice where the momentum is defined in 
the Brillouin zone
\begin{equation}
-\frac{\pi}{2a}\leq k_{\mu} < \frac{3\pi}{2a}.
\end{equation}
We assume that the operator $D$ is free of  species doubling,
which is proved for the explicit construction of $D$; in 
other words, the operator $D$ blows up rapidly 
($\sim \frac{1}{a}$) for small $a$ in the momentum region 
corresponding to species doublers. The contributions of doublers 
are eliminated by the regulator $f(x)$ in the above expression, 
since
\begin{equation}
tr\{\gamma_{5}f(\frac{(\gamma_{5}D)^{2}}{M^{2}})\}\sim
(\frac{1}{a})^{4}f(\frac{1}{(aM)^{2}})\rightarrow 0
\end{equation}
for $a\rightarrow 0$ if one chooses $f(x)=e^{-x}$, for example. 

We thus examine the above trace in the momentum range of the 
physical species
\begin{equation}
-\frac{\pi}{2a}\leq k_{\mu} < \frac{\pi}{2a}.
\end{equation}
We obtain the limiting $a\rightarrow 0$ expression
\begin{eqnarray}
&&\lim_{a\rightarrow 0}tr\{\gamma_{5}f(\frac{(\gamma_{5}D)^{2}}
{M^{2}})\}(x,x)\nonumber\\
&=& \lim_{a\rightarrow 0}tr \int_{-\frac{\pi}{2a}}^{\frac{\pi}
{2a}}\frac{d^{4}k}{(2\pi)^{4}}e^{-ikx}\gamma_{5}
f(\frac{(\gamma_{5}D)^{2}}{M^{2}})e^{ikx}\nonumber\\
&=&\lim_{L\rightarrow\infty}\lim_{a\rightarrow 0}tr 
\int_{-L}^{L}\frac{d^{4}k}{(2\pi)^{4}}e^{-ikx}\gamma_{5}
f(\frac{(\gamma_{5}D)^{2}}{M^{2}})e^{ikx}\nonumber\\
&=&\lim_{L\rightarrow\infty}tr \int_{-L}^{L}
\frac{d^{4}k}{(2\pi)^{4}}e^{-ikx}\gamma_{5}
f(\frac{(i\gamma_{5}\Dslash)^{2}}{M^{2}})e^{ikx}\nonumber\\
&\equiv&tr\{\gamma_{5}f(\frac{\Dslash^{2}}{M^{2}})\}
\end{eqnarray}
where  we first take the limit $a\rightarrow 0$ with fixed 
$k_{\mu}$ in $-L\leq k_{\mu} \leq L$, and then take the limit 
$L\rightarrow \infty$. This 
procedure is justified if the integral is well convergent.
\cite{fujikawa2} 
We 
also {\em assumed} that the 
operator $D$ satisfies  the following relation in the limit 
$a\rightarrow 0$
\begin{eqnarray}
De^{ikx}h(x) &\rightarrow& e^{ikx}(-\kslash+i\delslash
-g\Aslash)h(x)\nonumber\\
&=&i(\delslash+ig\Aslash)(e^{ikx}h(x))
\equiv i\Dslash(e^{ikx}h(x))
\end{eqnarray}
for any {\em fixed} $k_{\mu}$, ($-\frac{\pi}{2a}< k_{\mu}<
\frac{\pi}{2a}$), and a sufficiently smooth function $h(x)$. The 
function $h(x)$ corresponds to the gauge potential in our case, 
which in turn means that the gauge potential $A_{\mu}(x)$
is assumed to vary very little over the distances of the 
elementary lattice spacing. 

Our final expression (3.14) in the limit $M\rightarrow\infty$ 
reproduces the Pontryagin number in the continuum formulation  
(with $ \epsilon^{1234}=1$)\cite{fujikawa}
\begin{eqnarray}
\lim_{M\rightarrow\infty}tr \gamma_{5}f(\Dslash^{2}/M^{2})
&=&tr \gamma_{5}\frac{1}{2!}\{\frac{ig}{4}
[\gamma^{\mu},\gamma^{\nu}]F_{\mu\nu}\}^{2}
\int\frac{d^{4}k}{(2\pi)^{4}}f^{\prime\prime}(-k_{\mu}k^{\mu})
\nonumber\\
&&=\frac{g^{2}}{32\pi^{2}}tr \epsilon^{\mu\nu\alpha\beta}
F_{\mu\nu}F_{\alpha\beta}.
\end{eqnarray}
 
When one combines (3.7) and (3.16), one reproduces  the 
Atiyah-Singer index theorem (in continuum $R^{4}$ space)
.\cite{jackiw,atiyah} 
We  note that a local version of the index (anomaly) is valid 
for Abelian theory also.
The global index (3.7) as well as a local version of the index 
(3.8) are both independent of the regulator  $f(x)$  
provided\cite{fujikawa} 
\begin{equation}
f(0) =1, \ \ \ f(\infty)=0,\ \ \ f^{\prime}(x)x|_{x=0}=f^{\prime}
(x)x|_{x=\infty}=0. 
\end{equation}

We have thus established that the lattice index in (3.7) for any 
algebraic relation in (1.2) is related to the Pontryagin index 
in a smooth continuum limit as
\begin{equation}
n_{+}-n_{-}=\int d^{4}x\frac{g^{2}}{32\pi^{2}}tr
\epsilon^{\mu\nu\alpha\beta}F_{\mu\nu}F_{\alpha\beta}.
\end{equation} 
This shows  that  the instanton-related topological property is 
identical for all the algebraic relations in (1.2), and the 
Jacobian factor (3.6) in fact contains the 
correct chiral anomaly. (We are  implicitly assuming that the 
index (3.7) does not change in the process of taking a continuum 
limit.) 

A detailed perturbative analysis of chiral anomaly for the 
general operators with $k>0$ has been performed, and the 
above result has been confirmed.\cite{fujikawa2} Also a 
numerical study of 
the index relation has been performed: The numerical result 
indicates the consistency of our analyses.\cite{chiu2}  

\section{Explicit construction of the lattice Dirac operator 
for $k>1$}

We now comment on  an explicit construction of the lattice Dirac 
operator which satisfies the generalized algebraic relation
(1.2) with $k>0$.
We start with the conventional Wilson fermion operator $D_{W}$
defined by
\begin{eqnarray}
D_{W}(x,y)&\equiv&i\gamma^{\mu}C_{\mu}(x,y)+B(x,y)-
\frac{1}{a}m_{0}\delta_{x,y},\nonumber\\
C_{\mu}(x,y)&=&\frac{1}{2a}[\delta_{x+\hat{\mu} a,y}
U_{\mu}
(y)-\delta_{x,y+\hat{\mu} a}U^{\dagger}_{\mu}(x)],
\nonumber\\
B(x,y)&=&\frac{r}{2a}\sum_{\mu}[2\delta_{x,y}-
\delta_{y+\hat{\mu} a,x}U_{\mu}^{\dagger}(x)
-\delta_{y,x+\hat{\mu} a}U_{\mu}(y)],
\nonumber\\
U_{\mu}(y)&=& \exp [iagA_{\mu}(y)],
\end{eqnarray}
where we added a constant mass term to $D_{W}$ for later 
convenience. The parameter $r$ stands for the Wilson parameter.
Our matrix convention is that $\gamma^{\mu}$ are anti-hermitian, 
$(\gamma^{\mu})^{\dagger} = - \gamma^{\mu}$, and thus 
$\Cslash\equiv \gamma^{\mu}C_{\mu}(n,m)$ is hermitian
\begin{equation}
\Cslash^{\dagger} = \Cslash.
\end{equation}

The Dirac operator for a general value of $k$ is constructed 
by rewriting (1.2) as a set of relations
\begin{eqnarray}
&&H^{2k+1}\gamma_{5}+\gamma_{5}H^{2k+1}=2H^{2(2k+1)},\nonumber\\
&&H^{2}\gamma_{5}-\gamma_{5}H^{2}=0,
\end{eqnarray}
with $H=a\gamma_{5}D$. The second relation in (4.3) is shown by
using the defining relation (1.4), and the first of these 
relations (4.3) becomes
identical to the ordinary Ginsparg-Wilson relation (1.1) if one 
defines
$H_{(2k+1)}\equiv H^{2k+1}$. 
One can thus construct a solution to (4.3) by following the 
prescription used by Neuberger\cite{neuberger}
\begin{equation}
H_{(2k+1)}=\frac{1}{2}\gamma_{5}[1+D_{W}^{(2k+1)}\frac{1}
{\sqrt{(D_{W}^{(2k+1)})^{\dagger}D_{W}^{(2k+1)}}}]
\end{equation} 
where 
\begin{equation}
D_{W}^{(2k+1)}\equiv i(\Cslash)^{2k+1}+B^{2k+1}
-(\frac{m_{0}}{a})^{2k+1}
\end{equation}
The operator $H$ itself is then finally defined by (in the 
representation where $H_{(2k+1)}$ is diagonal)
\begin{equation}
H=(H_{(2k+1)})^{1/2k+1}
\end{equation}
in such a manner that the second relation of (4.3) is satisfied.
This condition (4.3) is shown to be satisfied 
 in the representation where $H_{(2k+1)}$ is 
diagonal.\cite{fujikawa2} 
 Also the conditions $0<m_{0}<2r=2$
and 
\begin{equation}
2m_{0}^{2k+1}=1
\end{equation}
ensure the absence of species doublers and a proper 
normalization of the Dirac operator $H$.

\section{Locality properties of general operators}
 
We have explained that the general operators for any finite
$k$ give rise to correct chiral anomaly and index relations in 
the (naive) continuum limit. This suggests that 
those operators are local for sufficiently smooth background
gauge field configurations. The locality of the standard overlap
operator with $k=0$ has been established by Hernandez, Jansen 
and L\"{u}scher,\cite{hernandez} and by 
Neuberger.\cite{neuberger2}   

As for the direct proof of locality of the operator $D$ for 
general $k$, one can show it for the 
vanishing gauge field by using the explicit solution
for the operator $H$ in momentum 
representation \cite{fujikawa2,chiu2}
\begin{eqnarray}
H(ap_{\mu})&=&\gamma_{5}(\frac{1}{2})^{\frac{k+1}{2k+1}}
(\frac{1}{\sqrt{H^{2}_{W}}})^{\frac{k+1}{2k+1}}
\{(\sqrt{H^{2}_{W}}+M_{k})^{\frac{k+1}{2k+1}}
-(\sqrt{H^{2}_{W}}-M_{k})^{\frac{k}{2k+1}}
\frac{\sslash}{a} \}\nonumber\\
&=&\gamma_{5}(\frac{1}{2})^{\frac{k+1}{2k+1}}
(\frac{1}{\sqrt{F_{(k)}}})^{\frac{k+1}{2k+1}}
\{(\sqrt{F_{(k)}}+\tilde{M}_{k})^{\frac{k+1}{2k+1}}
-(\sqrt{F_{(k)}}-\tilde{M}_{k})^{\frac{k}{2k+1}}
\sslash \}\nonumber\\
&&
\end{eqnarray}
where
\begin{eqnarray}
F_{(k)}&=&(s^{2})^{2k+1}+\tilde{M}_{k}^{2},\nonumber\\
\tilde{M}_{k}&=&[\sum_{\mu}(1-c_{\mu})]^{2k+1}
-m_{0}^{2k+1}
\end{eqnarray}
and
\begin{eqnarray}
&&s_{\mu}=\sin ap_{\mu}\nonumber\\
&&c_{\mu}=\cos ap_{\mu}\nonumber\\
&&\sslash=\gamma^{\mu}\sin ap_{\mu}.
\end{eqnarray}
For $k=0$, this operator is reduced to  Neuberger's overlap 
operator.\cite{neuberger} 
Here the inner product is defined to be $s^{2}\geq 0$.
This operator is shown to be free of species doublers for the 
 parameter $m_{0}$ within the range  $0<m_{0}<2$ when we set 
$r=1$, and $2m^{2k+1}_{0}=1$ gives a proper normalization
of $H$, namely, for an infinitesimal $p_{\mu}$, i.e.,
for $|ap_{\mu}|\ll 1$,
\begin{equation}
H\simeq-\gamma_{5}a\pslash(1+O(ap)^{2})
+\gamma_{5}(\gamma_{5}a\pslash)^{2k+2}
\end{equation}
to be consistent with $H=\gamma_{5}aD$; the last term in the 
rigth-hand side is the 
leading term of chiral symmetry breaking terms.

The locality of this explicit construction (5.1) is shown by 
studying
the analytic properties in the Brillouin zone.\cite{fujikawa2} 
It is 
important to recognize that this operator is not ultra-local but 
exponentially local;\cite{horvath} the operator $H(x,y)$ decays 
exponentially 
for large separation in coordinate representation
\begin{equation}
H(x,y)\sim \exp[-|x-y|/(2.5ka)].
\end{equation}

An explicit analysis of the locality of the operator $H_{(2k+1)}$
({\em not} $H$ itself)in the presence of gauge field, in 
particular, the locality 
domain for the gauge field strength $||F_{\mu\nu}||$ has been 
performed. The locality domain for $||F_{\mu\nu}||$ becomes 
smaller for larger $k$, 
but a definite non-zero domain has been 
established.\cite{fujikawa2} The 
remaing 
task is to show the locality domain of $||F_{\mu\nu}||$ for the 
operator $H=(H_{(2k+1)})^{1/(2k+1)}$. Due to the operation of 
taking the $(2k+1)$th root, an explicit analysis has not been 
performed yet, though a supporting argument has been 
given.\cite{fujikawa2}.

\section{Conclusion}
We have reported the recent investigation of topological
properties of a general class of lattice Dirac operators defined 
by the algebraic relation (1.2).
All these operators satisfy the index theorem and thus 
they are topologically proper. A precise proof of the locality 
of these general Dirac operators with fully dynamical gauge 
fields remains to be formulated. The operators with large $k$ 
is expected to exhibit infrared singularities in perturbative 
analyses as is 
suggested by the construction of $H_{(2k+1)}$ in (4.4), and 
thus the Wilsonian formulation of effective action, which is 
supposed to be free of infrared singularities, would be
essential.

Although we discussed only  4-dimensional theory, the 
recent developments in the treatment of lattice 
fermions\cite{niedermayer} may 
have some implications on 2-dimensional theory also, which is 
the main subject of this Symposium. In this respect, the fact 
that the lattice Dirac operators are not ultra-local but 
exponentially local\cite{hernandez} may be of some interest.
See Ref.\cite{balachandran} for a Ginsparg-Wilson construction
 on a 2-dimensional fuzzy sphere.
\\
 
\section*{Acknowledgments}
I thank Mo-lin Ge for the hospitality at Nankai
University.

\end{document}